\documentclass[conference]{IEEEtran}

\usepackage[utf8]{inputenc}
\usepackage[T1]{fontenc}
\usepackage{lmodern}
\usepackage{microtype}
\usepackage{cite}
\usepackage{amsmath,amssymb,amsfonts}
\usepackage{graphicx}
\usepackage{textcomp}
\usepackage{xcolor}
\usepackage{fancyvrb}
\usepackage{booktabs}
\usepackage{tikz}
\usetikzlibrary{arrows.meta,positioning,shapes.geometric}
\usepackage{url}
\usepackage{hyperref}
\hypersetup{colorlinks=true,linkcolor=black,citecolor=black,urlcolor=blue}
\DefineVerbatimEnvironment{code}{Verbatim}{fontsize=\footnotesize,fontfamily=cmtt,frame=single,framesep=2mm,xleftmargin=2mm,xrightmargin=2mm}

\newcommand{\VMG}{\mathit{ValidMsgGrammar}}

\newcommand{\copyrighttext}{%
  \footnotesize \textcopyright~2026 IEEE. Personal use of this material is permitted. Permission from IEEE must be obtained for all other uses, in any current or future media, including reprinting/republishing this material for advertising or promotional purposes, creating new collective works, for resale or redistribution to servers or lists, or reuse of any copyrighted component of this work in other works.}
\newcommand{\copyrightnotice}{%
  \begin{tikzpicture}[remember picture,overlay]
    \node[anchor=south,yshift=10pt] at (current page.south)
      {\parbox{\dimexpr0.92\textwidth\relax}{\copyrighttext}};
  \end{tikzpicture}%
}

\begin{document}

\title{CBCL: Safe Self-Extending Agent Communication}

\IEEEoverridecommandlockouts

\author{\IEEEauthorblockN{Hugo O'Connor}
\IEEEauthorblockA{Anuna Research\\
hugo@anuna.io}
\thanks{Accepted at the LangSec Workshop, 2026 IEEE Security and Privacy Workshops (SPW).}}

\maketitle
\copyrightnotice

\begin{abstract}
Agent communication languages (ACLs) enable heterogeneous agents to share knowledge and coordinate across diverse domains. This diversity demands extensibility, but expressive extension mechanisms can push the input language beyond the complexity classes where full validation is tractable. We present CBCL (Common Business Communication Language), an agent communication language that constrains all messages, including runtime language extensions, to the deterministic context-free language (DCFL) class. CBCL allows agents to define, transmit, and adopt domain-specific ``dialect'' extensions as first-class messages; three safety invariants (R1--R3), machine-checked in Lean~4 and enforced in a Rust reference implementation, prevent unbounded expansion, applying declared resource limits, and preserving core vocabulary. We formalize the language and its safety properties in Lean~4, implement a reference parser and dialect engine in Rust with property-based and differential tests, and extract a verified parser binary. Our results demonstrate that homoiconic protocol design, where extension definitions share the same representation as ordinary messages, can be made provably safe. As autonomous agents increasingly extend their own communication capabilities, formally bounding what they can express to each other is a precondition for oversight.
\end{abstract}


\section{Introduction}

Autonomous software agents increasingly coordinate through message-passing protocols to perform complex tasks: managing supply chains, orchestrating IoT device swarms, and mediating human--AI collaboration. The correctness and security of these systems depend critically on how agents parse and interpret the messages they exchange.

Language-theoretic security (LangSec), as articulated by Sassaman, Patterson, Bratus, and Locasto in \textit{Security applications of formal language theory}~\cite{sassaman2013}, shows that the computational complexity class of an input language determines the \emph{difficulty of validation} and the attack surface that parser implementations exhibit. Languages above the deterministic context-free class introduce ambiguity that makes parser equivalence undecidable~\cite{sassaman2013}, while Turing-complete inputs make validation itself undecidable; in both cases, ad-hoc validation cannot reliably prevent the emergence of ``weird machines''~\cite{bratus2011}, unintended computational artifacts that attackers exploit to achieve behavior beyond a system's intended functionality.

Contemporary agent communication presents a particularly acute instance of this problem. Early ACLs such as KQML~\cite{finin1994} and FIPA-ACL~\cite{fipa2002} defined core vocabularies of communicative acts (``performatives''). These languages are parseable and well-defined; KQML explicitly allowed communities to define new performatives by agreement, but without formal constraints on such extensions. In practice, extensibility was largely pushed into content languages and ontologies. Surveys of ACLs identify persistent interoperability failures. Heterogeneous agents remain confined to rigid contexts, and ad-hoc extensions proliferate incompatible dialects that are difficult to generalize without the original developers~\cite{kone2000state,chaibdraa2002trends}. In the terminology of Zambonelli et al.~\cite{zambonelli2011self}, traditional ACLs support only \emph{self-adaptation} (parameter tuning within a fixed structure), not \emph{self-expression} (structural evolution of the system itself). The modern alternative, using natural language or unrestricted JSON schemas as the communication substrate (as in MCP~\cite{mcp2024} and LLM-based agent frameworks), provides unlimited extensibility but creates an input language whose computational complexity is effectively unbounded. Determining whether an arbitrary natural language message will cause harmful behavior requires solving undecidable problems.

This paper presents CBCL, a language designed to be expressive, extensible, and tractable. CBCL provides a minimal core vocabulary (8 performatives) and a formal mechanism for agents to define, exchange, and adopt new domain-specific vocabularies (``dialects'') at runtime, without centralized coordination and without escaping the DCFL complexity class. CBCL achieves this through \emph{homoiconic self-extension}. Dialect definitions are themselves valid CBCL messages expressed in S-expression syntax, so they can be transmitted, verified, and installed using the same parsing machinery used for ordinary communication. Three safety constraints, verified both in a Lean~4 formalization and enforced in a Rust reference implementation, ensure that this self-extension is provably safe:

\begin{itemize}
\item \textbf{R1 (No Recursion):} Dialect templates are purely declarative pattern-template substitutions. Direct and mutual recursion are forbidden (no cyclic dependencies between performative templates). No iteration or reflection is permitted.
\item \textbf{R2 (Resource Bounds):} Every dialect declares static resource limits (nesting depth $\leq 64$, expansion size $\leq 8192$ characters, verification time $\leq 1000$\,ms) enforced at both installation and runtime.
\item \textbf{R3 (Core Preservation):} The eight core performatives cannot be redefined by any dialect, preserving protocol bootstrap invariants.
\end{itemize}

Together, these constraints support \emph{DCFL preservation}, meaning that the language recognized by a CBCL parser remains in DCFL regardless of how many dialects are installed. We give a formal argument in Section~\ref{sec:dcfl}; the full closure proof is mechanized in Lean~4 (theorem \texttt{dcfl\_preserved}).

CBCL is named after McCarthy's 1982 proposal for a ``Common Business Communication Language''~\cite{mccarthy1982} that would be ``open ended so that as programs improve, programs that can at first only order by stock numbers can later be programmed to inquire about specifications and prices\ldots'' Our work aspires to this vision while providing formal security guarantees.

\smallskip\noindent\textbf{Contributions.}
(1)~A protocol design methodology that achieves runtime self-extension while preserving DCFL membership, applicable to message-passing systems where DCFL expressiveness is acceptable.
(2)~A Lean~4 formalization covering parser correctness (soundness and completeness), safety constraint verification (R1--R3), pipeline totality for parse/validate/verify, and DCFL preservation under dialect installation.
(3)~A Rust reference implementation (\texttt{cbcl-rs}) with dialect verification, gossip propagation, C~FFI/WASM bindings, and fuzz targets; a verified parser binary extracted from the Lean proofs.
(4)~A draft IETF Internet-Draft specifying \texttt{application/cbcl}~\cite{oconnor2025}, with a proof-of-concept Nostr~\cite{nostr2020} transport binding (\texttt{cbcl-nostr}) for agent communication and Lightning Network micropayments over decentralized relays.

\section{Background and Threat Model}

\subsection{The Unbounded Attack Surface Problem}

Modern agent communication protocols have moved \emph{up} the Chomsky hierarchy~\cite{chomsky1956}, from the well-defined grammars of KQML and FIPA-ACL to unbounded complexity, without acknowledging the security consequences implied by formal language theory~\cite{sassaman2013,bratus2011} (Table~\ref{tab:protocols}).

\begin{table}[t]
\centering
\caption{Agent communication protocols classified by input language complexity.}
\label{tab:protocols}
\small
{\setlength{\tabcolsep}{4pt}
\begin{tabular}{@{}lcccc@{}}
\toprule
\textbf{Protocol} & \textbf{Complexity} & \textbf{Ext.} & \textbf{Verif.} & \textbf{Weird M.} \\
\midrule
KQML~\cite{finin1994} & CFL & Partial & Partial & Stack \\
FIPA-ACL~\cite{fipa2002} & CFL & Partial & Partial & Stack \\
MCP~\cite{mcp2024} & RE & Yes & No & Turing \\
LLM Agents & RE & Yes & No & Turing \\
\textbf{CBCL} & \textbf{DCFL} & \textbf{Yes} & \textbf{Yes} & \textbf{None\textsuperscript{\textdagger}} \\
\bottomrule
\end{tabular}
}
\vspace{1pt}

{\scriptsize ``Complexity'' refers to the effective computational power of the end-to-end message interpretation pipeline, not the surface syntax alone. KQML and FIPA-ACL use S-expression envelope syntax (DCFL), but neither specification constrains the content language, so the end-to-end complexity is at least CFL. MCP and LLM agent frameworks can induce Turing-complete behavior when tool invocation permits arbitrary code execution. CFL = context-free language (Type~2); DCFL = deterministic CFL; RE = recursively enumerable (Type~0). Ext.\@ = runtime self-extension of the protocol's vocabulary; ``Partial'' indicates informal extensibility without safety guarantees (KQML's open performative set, FIPA-ACL's \texttt{X-} parameters). Weird~M.\@ = weird machine class~\cite{bratus2011}. \textsuperscript{\textdagger}At the parser and template-expander layer; tool backends and agent action semantics are outside CBCL's scope.}
\end{table}

The threat model for CBCL assumes agents operating in open Byzantine environments where: (1)~\emph{untrusted peers} may send maliciously crafted messages designed to exploit parser differentials, trigger resource exhaustion, or inject unintended computation; (2)~\emph{dialect definitions} arrive from potentially adversarial sources and constitute executable extensions to the receiver's behavior; and (3)~\emph{no central authority} controls participant conduct or vocabulary evolution.

The security goal is to ensure that \emph{no message, including dialect definitions, can cause a conformant parser to enter an unanticipated state or perform unexpected computation}.

\subsection{Why DCFL Is the Right Complexity Class}

The choice of DCFL (rather than regular, general CFG, or context-sensitive) is deliberate:

\begin{itemize}
\item \textbf{Regular languages} cannot express the nested structure of agent messages (envelopes wrapping messages, dialects scoping inner messages).
\item \textbf{General CFL} risks parser differentials: ambiguous grammars admit multiple parse trees, and attackers can exploit divergent interpretations~\cite{sassaman2013,kaminsky2010}. Every DCFL has an unambiguous grammar, making unambiguity a class-level property rather than a per-grammar proof obligation~\cite{hopcroft2006}. Parser equivalence is therefore preserved automatically under language evolution.
\item \textbf{DCFL} is the minimal class supporting nested structure with \emph{parser equivalence}. Residual divergence from serialization or encoding is mitigated through canonical serialization (Section~\ref{sec:canonical}).
\item \textbf{Context-sensitive and beyond.} Context-sensitive membership is PSPACE-complete~\cite{hopcroft2006}; Type~0 is undecidable. Both exceed the recognizer complexity that LangSec considers tractable for full input validation~\cite{sassaman2013}.
\end{itemize}

S-expression syntax is a natural fit: the grammar is LR(1)-parseable (indeed LL(1)) and nested parenthesized lists directly mirror the pushdown automaton's stack operations.

\section{CBCL Language Design}

We use the following terminology throughout. A \emph{performative} is a named message type (e.g., \texttt{tell}, \texttt{ask}). A \emph{dialect} is a named, versioned collection of performative definitions that extends CBCL's core vocabulary. Each dialect performative has a \emph{template}: the pattern-template substitution rule that maps dialect invocations to core CBCL messages.

\subsection{Core Grammar}

CBCL messages are S-expressions conforming to an ABNF grammar (fully specified in the IETF draft~\cite{oconnor2025}). The top-level rules are:

\begin{code}
cbcl-message = simple-message / meta-message
             / lang-message  / wrapped-message
simple-message = "(" performative SP recipient
                 [SP content] *(SP parameter) ")"
SP             = " "   ; single ASCII space (0x20)
performative   = "tell" / "ask" / "reply" / "hello"
               / "bye" / "ok" / "error" / "cancel"
lang-message   = "(" "lang" SP dialect-name
                 SP cbcl-message ")"
s-expr         = "(" *( s-expr / atom ) ")"
atom           = identifier / string / number
               / boolean / symbol
\end{code}

\noindent The four message categories are:

\smallskip\noindent\textbf{Simple messages} use one of eight core performatives:

\begin{code}
(tell @bob "The meeting is at 3pm")
(ask @alice "Status of task-42?" :thread conv-17)
(reply @alice "75% complete" :in-reply-to msg-9874)
(ok @bob)  (error @sender "parse failure")
(hello @bob)  (bye @bob)
\end{code}

\noindent Colon-prefixed identifiers such as \texttt{:thread} and \texttt{:in-reply-to} are \emph{keyword parameters}: named optional arguments in the style of Common Lisp keyword symbols. They are syntactically atoms (the \texttt{symbol} production in the grammar) and carry no special semantics beyond serving as self-quoting parameter labels.

\smallskip\noindent\textbf{Meta messages} operate on dialects---defining, querying, and teaching:

\begin{code}
(meta (define logistics (cbcl) @consortium
  (:resource-requirements
    ((max-depth 16)
     (max-expansion-size 4096)
     (verification-time 1000)))
  (extend track-shipment (pkg-id &key route priority)
    (tell @tracking-svc
      (shipment-request :package pkg-id
        :route route
        :priority (or priority "normal"))
      :domain logistics))
  (:examples
    (track-shipment "PKG-42" :route "A->B")
    (tell @tracking-svc
      (shipment-request :package "PKG-42"
        :route "A->B" :priority "normal")
      :domain logistics))))
\end{code}

\noindent The optional \texttt{:examples} clause carries semantic meaning: each input/output pair illustrates what the dialect author \emph{intends} a performative to do in a concrete scenario. Because CBCL cannot formally constrain meaning (Section~\ref{sec:semantic}), examples serve as the primary mechanism for communicating intended semantics between agents. A receiving agent can additionally evaluate the examples against its own expander to verify mechanical agreement before claiming dialect support.

\noindent A \texttt{query} meta message asks a peer whether it supports a dialect; a \texttt{teach} meta message transmits a previously defined dialect to another agent:

\begin{code}
(meta (query logistics @bob))
(meta (teach logistics @alice
  <dialect-definition>))
\end{code}

\noindent The \texttt{teach} payload is the same S-expression used in \texttt{define}; the receiving agent parses, verifies (R1--R3), and installs it using its existing CBCL infrastructure.

\smallskip\noindent\textbf{Language-scoped messages} invoke dialect performatives. Dialect invocations \emph{must} appear inside a \texttt{lang} wrapper that names the dialect; bare dialect performatives at top level are syntactically invalid. This scoping rule is what makes the DCFL preservation argument (Section~\ref{sec:dcfl}) straightforward: the \texttt{lang} tag serves as a deterministic dispatch token.

\begin{code}
(lang logistics
  (track-shipment "PKG-123" :route "A->B"))
\end{code}

\noindent This expands via pattern-template substitution to:

\begin{code}
(tell @tracking-svc
  (shipment-request :package "PKG-123"
    :route "A->B" :priority "normal")
  :domain logistics)
\end{code}

\noindent\textbf{Wrapped messages} provide metadata (envelopes), integrity (signatures), and resource governance:

\begin{code}
(with-limits :timeout 100 :max-depth 10
  (ask @reasoner "Compute optimal path"))
\end{code}

\subsection{Homoiconic Self-Extension}

CBCL's dialect definitions are themselves valid CBCL messages, eliminating the need for a separate meta-language parser (a second attack surface) or out-of-band registry. Inspired by Racket's \texttt{\#lang} mechanism~\cite{flatt2012}, a single DCFL recognizer handles both ordinary communication and language extension; the DCFL preservation argument (Section~\ref{sec:dcfl}) explains why installing a dialect does not increase the recognizer's computational power.

\subsection{Template Expansion Semantics}

Dialect performatives expand through a deliberately restricted template language supporting four forms: (1)~literal templates with parameter placeholders, (2)~substitution of parameter references with argument values, (3)~bounded conditionals \texttt{(cond ((= param value) template) ...)} with no recursion, and (4)~sequences of template expressions.

This template language is \emph{not Turing-complete}. It supports only flat pattern matching, finite conditional branching, and one-shot substitution. Critically, expansion is \emph{single-pass}: the expanded result is not fed back through the expander. This ensures that template evaluation terminates in time linear in the template size plus substituted argument size, and is bounded by declared expansion limits.

\subsection{Message Processing Lifecycle}

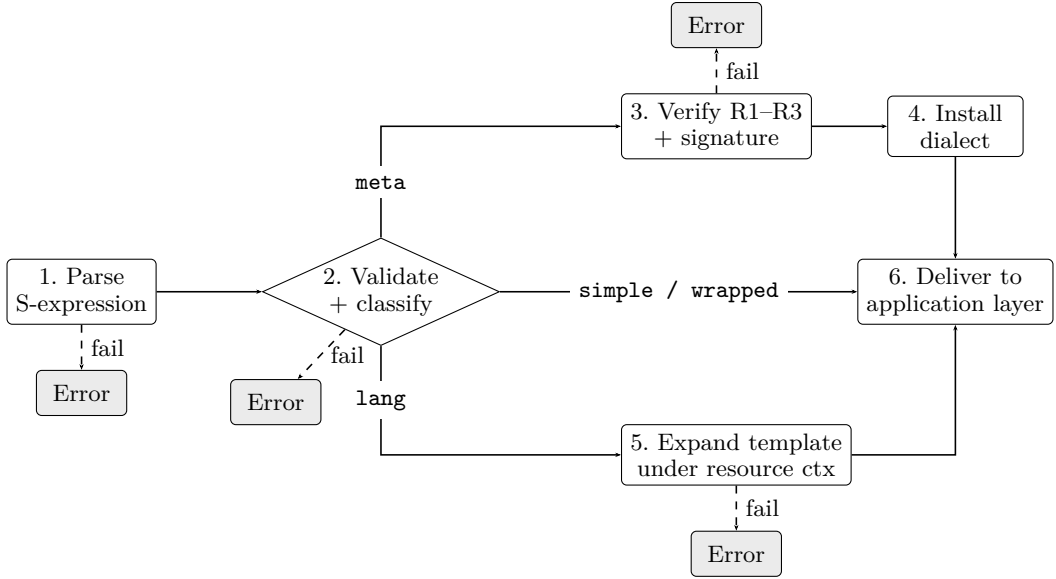
\begin{figure*}[t]
\centering
\begin{tikzpicture}[
  node distance=5mm and 6mm,
  every node/.style={font=\small},
  proc/.style={rectangle, draw, rounded corners=2pt,
               minimum width=18mm, minimum height=7mm,
               fill=white, align=center},
  decision/.style={diamond, draw, aspect=2.2,
                   minimum width=16mm, minimum height=6mm,
                   fill=white, align=center, inner sep=1pt},
  err/.style={rectangle, draw, rounded corners=2pt,
              minimum width=12mm, minimum height=6mm,
              fill=black!8, align=center},
  arr/.style={-{Stealth[length=2.2pt]}, semithick},
]
\node[proc] (parse) {1.~Parse\\S-expression};
\node[decision, right=14mm of parse] (type) {2.~Validate\\+ classify};

\node[proc, above right=14mm and 24mm of type] (verify) {3.~Verify R1--R3\\+ signature};
\node[proc, right=10mm of verify] (install) {4.~Install\\dialect};

\node[proc, below right=14mm and 24mm of type] (expand) {5.~Expand template\\under resource ctx};

\node[proc] at (install |- type) (deliver) {6.~Deliver to\\application layer};

\node[err, below=6mm of parse] (e1) {Error};
\node[err, below left=8mm and 0mm of type] (e2) {Error};
\node[err, above=6mm of verify] (e3) {Error};
\node[err, below=6mm of expand] (e4) {Error};

\draw[arr] (parse) -- (type);

\draw[arr] (type) |- node[fill=white, pos=0.25] {\texttt{meta}} (verify);
\draw[arr] (type) -- node[fill=white] {\texttt{simple / wrapped}} (deliver);
\draw[arr] (type) |- node[fill=white, pos=0.25] {\texttt{lang}} (expand);

\draw[arr] (verify) -- (install);
\draw[arr] (install) -- (deliver);
\draw[arr] (expand) -| (deliver);

\draw[arr, dashed] (parse) -- node[right] {fail} (e1);
\draw[arr, dashed] (type) -- node[right] {fail} (e2);
\draw[arr, dashed] (verify) -- node[right] {fail} (e3);
\draw[arr, dashed] (expand) -- node[right] {fail} (e4);

\end{tikzpicture}
\caption{Message processing lifecycle. Steps 1--2 apply to all messages; the upper branch handles dialect definitions (\texttt{meta}), the lower branch handles dialect invocations (\texttt{lang}), and simple or wrapped messages proceed directly to delivery. Any step may reject with a well-typed error (Theorem~12).}
\label{fig:lifecycle}
\end{figure*}

Figure~\ref{fig:lifecycle} shows the processing pipeline that every incoming message traverses. Steps 1--2 (parse, validate~+~classify) apply to all messages. The dialect-definition path (steps 3--4) applies only to \texttt{meta~define} messages; the expansion path (step~5) applies only to \texttt{lang} invocations. The pipeline is total: every input produces either a well-typed result or a well-typed error (Theorem~12).

\section{Safety Invariants and Formal Verification}

\subsection{Lean 4 Formalization Overview}

We formalize CBCL in Lean~4 as a library (\texttt{LeanCbcl}) comprising modules for S-expressions, messages, parsing, serialization, pattern matching, dialect parsing, template expansion, deterministic-union reasoning, dialect constraints (R1--R3), and an end-to-end pipeline. The formalization comprises approximately 5,400 lines of Lean across 16 modules and yields 176 machine-checked theorems, with zero \texttt{sorry} gaps and no custom axioms. A verified parser binary (\texttt{cbcl-parse}) is extracted from the proofs.

\subsection{Parser Correctness}

The S-expression parser uses recursive descent with a fuel parameter for termination. The fuel is set to $|s| \times 4 + 1$, where $|s|$ is the input length in characters. The multiplier~4 reflects the worst-case fuel consumption per input character: a single character can trigger up to four recursive steps (entering a list, attempting to parse an element, consuming the closing delimiter, and returning the result). The $+1$ ensures termination on empty input. This bound is sufficient for all inputs; the Lean proof of Theorem~1 confirms that \texttt{parseSExpr} never exhausts fuel. The fuel parameter is retained in the extracted \texttt{cbcl-parse} binary. The fuel value is computed at runtime from the input length, so the executable faithfully mirrors the verified model with no gap between proof and implementation. The top-level parser also rejects trailing input and unterminated strings.

\smallskip
\noindent\textbf{Theorem 1} (Termination). \textit{For all inputs $s$, $\mathit{parseSExpr}(s)$ terminates and produces either a valid \textnormal{\texttt{SExpr}} or a parse error.}

\smallskip
A universal well-formedness theorem (\texttt{allSExpr\-\_well\-Formed}) additionally proves by structural induction that \emph{every} \texttt{SExpr} value satisfies the inductive \texttt{Well\-Formed\-SExpr} predicate, ensuring the parser can only produce structurally valid output.

\smallskip
In these theorems, $e$ is an \texttt{SExpr} (the S-expression parse tree/AST) and $m$ is a \texttt{Message} (the semantic classification into simple, meta, dialect, or wrapped). The predicate $\VMG(e, m)$ is an inductive relation encoding which S-expressions constitute valid CBCL messages: it has four constructors corresponding to simple messages (with one of the eight core performatives), \texttt{meta} messages, \texttt{lang}-scoped dialect invocations, and wrapped (envelope/signed/with-limits) messages. Because the grammar is DCFL, each valid input admits exactly one parse tree, so the \texttt{Message} returned by the parser is unique.

\smallskip
\noindent\textbf{Theorem 2} (Soundness). \textit{If $\mathit{parseMessage}(e) = \mathit{some}\;m$, then $\VMG(e, m)$.}

\smallskip
\noindent\textbf{Theorem 3} (Completeness). \textit{If $\VMG(e, m)$ holds, then $\mathit{parseMessage}(e) = \mathit{some}\;m$.}

\smallskip
Soundness and completeness together establish that the parser recognizes \emph{exactly} the message grammar encoded in Lean (via \texttt{ValidMessageGrammar}), no more, no less. This eliminates the possibility of parser differential attacks between the formal specification and the implementation, a core LangSec concern.

We additionally verify 10 concrete round-trip properties ($\mathit{parse}(\mathit{serialize}(e)) = \mathit{Ok}(e)$) covering the seven atom types (symbol, integer, negative integer, boolean true/false, string, keyword), empty lists, flat lists, and nested lists. A \texttt{SafeSymbol} predicate formally characterizes which symbol names survive round-tripping (non-empty, no delimiters, not a boolean or integer literal); the general theorem \texttt{roundtrip\_safeSymbol} proves round-tripping for all safe symbols via char-level parser reasoning.

\subsection{R1: No Recursion}

Constraint R1 ensures that dialect performative templates contain no direct or mutual recursion, preventing unbounded expansion. The Lean~4 formalization covers both: Theorems~4--5 establish soundness and completeness for \emph{self-reference detection} (a performative name cannot appear in its own template), while Theorem~7 proves soundness of \emph{mutual recursion prevention} via a transitive-closure dependency predicate ($\mathit{DependsClosure}$) that detects cycles of any length in the performative dependency graph.

\smallskip
\noindent\textbf{Theorem 4} (R1 Soundness). \textit{If $\mathit{containsSelfRef}(\mathit{name}, t) = \mathit{false}$, then $\mathit{name}$ does not occur anywhere in $t$.}

\smallskip
\noindent\textbf{Theorem 5} (R1 Completeness). \textit{If $\mathit{name}$ occurs in $t$, then $\mathit{containsSelfRef}(\mathit{name}, t) = \mathit{true}$.}

\smallskip
\noindent\textbf{Theorem 6.} \textit{The base dialect (8 core performatives) is R1-valid.}

\smallskip
\noindent\textbf{Theorem 7} (Mutual Recursion Soundness). \textit{If $\mathit{verifyR1NoMutualRecursion}(\delta) = \mathit{true}$, then no performative in $\delta$ participates in a dependency cycle.}

\smallskip
The mutual recursion verifier uses a computable DFS-based cycle detector (\texttt{checkNoCycles}) with full soundness and completeness proofs against a semantic \texttt{mutualRecursion} specification: no false negatives and no false positives. The Rust reference implementation mirrors the computable version.

\subsection{R2: Resource Bounds}

Constraint R2 enforces static resource limits ($\mathit{maxDepth} \leq 64$, $\mathit{maxExpansionSize} \leq 8192$ characters, $\mathit{verificationTime} \leq 1000$\,ms). The Lean~4 model uses a fuel parameter: $\mathit{boundedEvalFull}$ decrements a natural-number counter at each expansion step, returning $\mathit{none}$ when fuel reaches zero. The Rust implementation mirrors this for depth and size, adding a wall-clock timeout as defense-in-depth.

\smallskip
\noindent\textbf{Theorem 8} (Bounded Evaluation). \textit{$\mathit{boundedEvalFull}(\mathit{fuel}, e, rs)$ always terminates, producing either a result or $\mathit{none}$ (resource exhaustion).} A depth-bounded invariant additionally rules out depth creep across nested expansions.

\smallskip
\noindent\textbf{Theorem 9.} \textit{If $\mathit{verifyR2}(\delta) = \mathit{true}$, then $\delta$'s declared bounds are within system limits.}

\subsection{R3: Core Preservation}

Constraint R3 prevents dialects from redefining the eight core performatives.

\smallskip
\noindent\textbf{Theorem 10} (Core Preservation). \textit{If $\mathit{verifyR3}(\delta) = \mathit{true}$, then for all core performative names $n$, $\delta$ does not define $n$.}

\smallskip
\noindent\textbf{Theorem 11} (Installation Safety). \textit{Installing an R3-verified dialect into a well-formed agent preserves well-formedness.}

\subsection{Pipeline Totality}

The Lean pipeline from parse $\to$ validate $\to$ verify constraints (for \texttt{define}/\texttt{define-dialect} meta messages) is formalized as $\mathit{runPipeline} : \mathit{String} \to \mathit{PipelineResult}$.

\smallskip
\noindent\textbf{Theorem 12} (Pipeline Totality). \textit{For all inputs $s$, $\mathit{runPipeline}(s)$ returns a $\mathit{PipelineResult}$ and never diverges.}

\smallskip
\noindent\textbf{Theorem 13} (Pipeline Soundness). \textit{If $\mathit{runPipeline}(s) = \mathit{success}(m)$, then $s$ parsed successfully, $m$ satisfies $\VMG$, and $m$ passes validation. If $m$ is a \texttt{define}/\texttt{define-dialect} meta message, its dialect definition passes R1--R3.}

\subsection{DCFL Preservation Argument}\label{sec:dcfl}

The central security property is that installing dialects does not push the recognized language beyond DCFL. We establish this through three steps:

\begin{enumerate}
\item \textbf{Base case.} The core CBCL grammar is LR(1)-parseable (indeed LL(1)), hence DCFL.
\item \textbf{Extension step.} A dialect adds finitely many \texttt{extend} clauses that introduce new performative heads. All dialect invocations must appear inside a \texttt{(lang \textit{name} ...)} wrapper. DCFLs are not closed under union in general~\cite{ginsburg1966}; CBCL avoids this because the \texttt{lang} keyword followed by a unique dialect name forms a prefix-free partition: a deterministic pushdown automaton (DPDA)~\cite{hopcroft2006} dispatches to the appropriate subgrammar by consuming the \texttt{lang} tag and dialect name, with no additional lookahead beyond LR(1) required. R1 and R2 ensure template expansion terminates and is resource-bounded, but do not increase parsing power.
\item \textbf{Inductive closure.} Since each installation preserves DCFL membership, and the base grammar is DCFL, the language remains DCFL after any finite sequence of dialect installations.
\end{enumerate}

This argument is fully mechanized in Lean~4. Theorems 2--3 establish parser correctness, 4--7 verify R1, 8--9 verify R2, and 10--11 verify R3. The deterministic-union construction is proven by \texttt{agentDetParser\_agrees}, which shows that the combined DPDA agrees with the boolean decider for agents with unique dialect names (\texttt{namesUnique}). The final theorem \texttt{dcfl\_preserved} composes these results: installing a fresh R3-verified dialect preserves DCFL membership.

\section{Reference Implementation}

\subsection{Architecture}

The reference implementation is written in Rust ({\raise.17ex\hbox{$\scriptstyle\sim$}}11,000 lines of library code) as a Cargo workspace of five crates (Table~\ref{tab:modules}). The core library (\texttt{cbcl-core}) forbids unsafe code and uses only \texttt{alloc}, making it suitable for \texttt{no\_std} environments. An earlier GNU Guile Scheme prototype informed the design but is superseded by the Rust implementation.

\begin{table}[t]
\centering
\caption{Reference implementation crates (Rust).}
\label{tab:modules}
\footnotesize
\begin{tabular}{@{}l@{\hspace{4pt}}r@{\hspace{6pt}}l@{}}
\toprule
\textbf{Crate} & \textbf{LOC} & \textbf{Responsibility} \\
\midrule
\texttt{cbcl-core} & 7{,}800 & Agent, evaluator, R1--R3, gossip, canonical \\
\texttt{cbcl-parser} & 1{,}900 & Recursive-descent S-expr \& message parser \\
\texttt{cbcl-cli} & 430 & Parse, verify, agent REPL, gossip simulation \\
\texttt{cbcl-ffi} & 430 & C~FFI bindings via \texttt{cbindgen} \\
\texttt{cbcl-wasm} & 490 & WebAssembly bindings (optional JS interop) \\
\bottomrule
\end{tabular}
\end{table}

\subsection{Dialect Verification Engine}

Upon receiving a dialect definition, the implementation executes: (1)~\textbf{R1 check}: walk each template AST for self-references and reject if found; additionally reject mutual recursion by detecting cycles in the performative dependency graph; (2)~\textbf{R2 check}: verify declared bounds are within system limits; (3)~\textbf{R3 check}: reject if any \texttt{extend} clause redefines a core performative; (4)~\textbf{Integrity check} (optional): if a cryptographic signer is registered (via a \texttt{Signer} trait), verify the signature and content hash. The protocol is algorithm-agnostic; the signature algorithm is identified by the dialect's metadata, allowing deployments to use Ed25519, ECDSA, or post-quantum schemes without protocol changes. Only after all checks pass is the dialect installed. Verification runs in $O(|\delta|^2)$ time in the size of the dialect definition.

\subsection{Runtime Resource Enforcement}

A \texttt{ResourceContext} struct tracks depth, cumulative expansion size, and elapsed wall-clock time during message evaluation. Three checks enforce bounds: \texttt{enter\_depth} increments the depth counter and returns a resource-limit error if \texttt{max\_depth} is exceeded; \texttt{check\_expansion\_limit} accumulates expansion size; and \texttt{check\_time\_limit} compares elapsed time against the timeout. These checks are invoked at each step of template expansion, ensuring that even adversarially crafted templates cannot exceed their declared resource bounds.

\subsection{Epidemic Dialect Propagation}

Dialects propagate between agents via an epidemic (gossip) protocol. In a fully-connected topology, classic epidemic dissemination analysis~\cite{demers1987} yields convergence to full coverage in $O(\log n)$ rounds; CBCL builds on this result. With transmission probability $p = 0.8$, empirical testing with networks of 10--500 agents matches the $O(\log n)$ bound; a 100-agent network achieves full coverage in $\sim$3 rounds. Receiving agents independently verify dialect constraints (R1--R3) and cryptographic signatures before installation.

\subsection{Canonical Serialization}\label{sec:canonical}

For cryptographic operations (signing, hashing), CBCL uses the canonical form of S-expressions specified in RFC~9804~\cite{rfc9804}. This encoding eliminates whitespace variation and uses length-prefixed verbatim strings, ensuring that structurally identical messages produce identical byte representations, a prerequisite for deterministic signature verification.

\section{Evaluation}

\subsection{Verification Coverage}

The Lean~4 formalization covers parsing, validation, dialect parsing/verification for meta definitions, and a fuel-bounded evaluation model. Table~\ref{tab:theorems} reports counts for the core safety modules.

\begin{table}[t]
\centering
\caption{Lean 4 verification coverage.}
\label{tab:theorems}
\footnotesize
\begin{tabular}{@{}lrl@{}}
\toprule
\textbf{Component} & \textbf{Thm.} & \textbf{Approach} \\
\midrule
S-Expression types & 2 & Structural induction \\
Parser & 26 & Fuel-based, well-formedness \\
Serializer & 51 & Round-trip, \texttt{SafeSymbol} \\
Message parser & 9 & Soundness + completeness \\
R1 (no recursion) & 13 & DFS sound.\ + complete.\ \\
R2 (resource bounds) & 6 & Depth-bounded evaluation \\
R3 (core preservation) & 4 & Redefinition check \\
Pipeline & 6 & Totality + composition \\
DCFL preservation & 50 & Det.\ union + det.\ parser \\
Agent \& dialect model & 9 & Installation + matching \\
\midrule
\textbf{Total} & \textbf{176} & \textbf{All machine-checked, zero \texttt{sorry}} \\
\bottomrule
\end{tabular}
\end{table}

The verified parser binary \texttt{cbcl-parse} is extracted from the Lean proofs and runs 17 built-in test cases covering all message types, including negative cases for trailing input and unterminated strings.

The Rust implementation includes property-based tests (using \texttt{proptest}), differential tests that cross-check the Rust parser against the Lean-extracted binary, and five \texttt{cargo-fuzz} targets covering the S-expression parser, message parser, dialect parser, constraint verification, and template expansion. Five example dialects (37 performatives total) have been verified against the implementation (R1--R3 pass): precision agriculture (11 performatives), AI planning (7), cross-chain asset transfer (7), artifact management (7), and email use (5).

\subsection{Security Testing}

We validate CBCL's security properties against concrete attack scenarios:

\smallskip\noindent\textbf{Recursive expansion attack.} A malicious dialect containing \texttt{(extend bomb (x) (bomb (bomb x)))} is rejected at R1 verification; the self-reference is detected before installation. Mutual recursion (e.g., \texttt{ping} expands to \texttt{pong}, \texttt{pong} expands to \texttt{ping}) is also rejected by the implementation.

\smallskip\noindent\textbf{Depth bomb.} A deeply nested message \texttt{(lang d1 (lang d2 (lang d3 ...)))} exceeding \texttt{max-depth} triggers resource exhaustion at the enforced limit, returning an error rather than consuming unbounded stack.

\smallskip\noindent\textbf{Core redefinition.} A dialect attempting \texttt{(extend tell (x) (drop-table x))} is rejected at R3 verification; \texttt{tell} is a core performative and cannot be redefined.

\smallskip\noindent\textbf{Expansion bomb.} A template producing output larger than \texttt{max-expansion-size} is halted mid-expansion when the cumulative size check fires.

\smallskip
In all cases, the parser returns a well-typed error value rather than entering an undefined state. This is the practical consequence of DCFL preservation: the parser is deterministic and total, always terminating with a definite accept or reject.

\subsection{Performance}

Table~\ref{tab:perf} reports Criterion benchmark results for the Rust implementation on an Apple~M4 (macOS~15). The end-to-end pipeline (parse~$\to$ validate~$\to$ verify) processes a simple \texttt{tell} message in under 400\,ns and a \texttt{meta~define} (including R1--R3 verification) in $\sim$1\,$\mu$s. Template expansion completes in 120--350\,ns depending on template complexity. Constraint verification is dominated by R1 (dependency-graph cycle detection, $\sim$1.8\,$\mu$s for a full dialect); R2 and R3 checks complete in under 25\,ns each. Gossip simulation of a 100-agent fully-connected network converges in $\sim$2.6\,ms. These numbers indicate that CBCL's safety checks impose negligible overhead relative to typical network round-trip times.

\begin{table}[t]
\centering
\caption{Benchmark results (Rust, Apple~M4).}
\label{tab:perf}
\footnotesize
\begin{tabular}{@{}lr@{}}
\toprule
\textbf{Operation} & \textbf{Median time} \\
\midrule
Parse simple \texttt{tell} message & 78\,ns \\
Parse \texttt{meta define} & 125\,ns \\
Parse \texttt{lang} dialect invocation & 130\,ns \\
Pipeline: simple \texttt{tell} (end-to-end) & 386\,ns \\
Pipeline: \texttt{meta define} (end-to-end) & 1.03\,$\mu$s \\
Template expansion (conditional) & 215\,ns \\
R1 verify full dialect & 1.83\,$\mu$s \\
R2 bounded eval (nested) & 185\,ns \\
R3 verify dialect & 21\,ns \\
Gossip: 100-agent convergence & 2.6\,ms \\
\bottomrule
\end{tabular}
\end{table}

\section{Discussion}

\subsection{Relation to LangSec Principles}

\noindent\textbf{Verifiable parsers.} The Lean~4 formalization produces a parser with machine-checked soundness and completeness theorems. The extracted binary constitutes a verified recognizer for the CBCL language.

\smallskip\noindent\textbf{Parser equivalence.} Because CBCL specifies a single unambiguous grammar that is LR(1)-parseable (and indeed LL(1)), all conformant parsers produce identical parse trees for every input. This eliminates the grammar-level parser differential vulnerabilities that arise when different implementations interpret the same message differently, a problem endemic to protocols with ambiguous grammars or natural language content. Canonical serialization (Section~\ref{sec:canonical}) further reduces---but cannot fully eliminate---implementation-level divergence in areas such as Unicode handling.

\smallskip\noindent\textbf{Explicit computational complexity.} Every dialect declares its resource requirements, and evaluation is guaranteed to terminate within those bounds. An implementation can decide \emph{before evaluation} whether a dialect's demands are acceptable, yielding explicit worst-case resource budgets. This transforms resource governance from a runtime emergency into a design-time contract.

\smallskip\noindent\textbf{No parser- or expander-induced weird machines.} The restriction to declarative pattern-template expansion with bounded resources means that the CBCL parser and template expander introduce no unintended computational artifacts. The template language is not Turing-complete; its semantics are fully determined by a finite substitution table. Weird machines could still arise in layers outside CBCL's scope (e.g., tool backends or agent action semantics); CBCL's guarantee is that the message recognition and expansion pipeline itself is free of them.

\subsection{Comparison with Contemporary Agent Protocols}

The Model Context Protocol (MCP)~\cite{mcp2024} and similar LLM agent frameworks use JSON-RPC with unrestricted schemas. While practical, this design creates an input language whose computational complexity depends on the tools agents invoke, potentially reaching Turing-completeness when they can execute arbitrary code in response to messages. CBCL demonstrates that meaningful extensibility does not require unbounded input complexity.

KQML~\cite{finin1994} and FIPA-ACL~\cite{fipa2002} have parseable grammars and limited extensibility (KQML's performative set was explicitly open; FIPA-ACL supported user-defined message parameters), but extensions were unconstrained by formal safety guarantees and required out-of-band agreement. CBCL inherits their grammar tractability while adding formally safe runtime evolution.

\subsection{Implementation Strategy}

The Lean~4 formalization and the Rust implementation serve complementary roles. Lean provides machine-checked proofs of the safety properties (Theorems~1--13) and yields an extracted parser binary (\texttt{cbcl-parse}) that is correct by construction. However, extracted code optimizes for proof structure, not for runtime performance or integration. The Rust implementation (\texttt{cbcl-rs}) provides a production-quality library with $O(n)$ hand-rolled parsing, C~FFI and WebAssembly bindings, property-based and differential testing, and fuzz targets---capabilities that are essential for adoption but outside the scope of formal verification. The two implementations are cross-validated: the Rust parser's test suite includes differential tests against the Lean-extracted binary, ensuring that both accept and reject the same inputs.

\subsection{Toward Semantic Guidance}\label{sec:semantic}

CBCL guarantees syntactic safety but not semantic agreement. Several mechanisms could narrow this gap, with different complexity implications. \emph{Examples} (supported via \texttt{:examples} in dialect definitions) and \emph{typed parameter declarations} add no formal complexity and remain within DCFL. \emph{Interaction protocol state machines} (e.g., after \texttt{ask}, expect \texttt{reply}) are regular, well below DCFL. \emph{Pre/postconditions} and \emph{ontological commitment}~\cite{gruber1993translation} risk crossing the complexity boundary CBCL is designed to enforce, and would need to be scoped carefully or relegated to an application layer above the protocol. The \texttt{:examples} clause offers a machine-checkable anchor: a receiving agent \textsc{should} expand each example input through its own expander and compare the result against the expected output before claiming dialect support. This catches the most common interoperability failures (template misconfiguration, parameter-type disagreement) before they manifest at runtime, though it cannot detect deeper semantic divergence. Following Singh~\cite{singh1998}, CBCL treats dialect installation as a public commitment to the dialect's declared semantics; semantic divergence is thus a breach of commitment, observable and accountable, rather than a hidden failure of shared mental state. We are investigating \emph{structural contracts} that extend CBCL's syntactic safety toward protocol correctness while remaining within DCFL and requiring no coordination. Protocol constraints are expressed as causal dependency graphs over performative types; each message carries an explicit causal reference (a content hash of its predecessor's canonical serialisation), forming a Merkle DAG that is cryptographically tamper-evident and independently verifiable. Verification is a monotonic predicate on the append-only message store, coordination-free by the CALM theorem~\cite{hellerstein2020}. Message shape constraints are expressed as visibly pushdown languages~\cite{alur2004} over expanded S-expressions, exploiting the decidable inclusion checking that VPLs provide. The full design is implemented in \texttt{cbcl-rs} and will be described in a forthcoming report.

\subsection{Limitations}

\textbf{Expressiveness ceiling.} DCFL restriction means some patterns cannot be expressed in dialect templates. Concrete examples of inexpressible patterns include: (1)~recursive data transformations such as flattening a nested list of arbitrary depth; (2)~cross-field references where one parameter's value depends on another parameter's content (e.g., a checksum computed over other fields); (3)~iteration or aggregation over variable-length collections (e.g., summing line items in an invoice); and (4)~context-sensitive validation such as ensuring a reply's thread ID matches an earlier message. In Standish's extensibility taxonomy~\cite{standish1975}, CBCL restricts itself to \emph{paraphrastic} extension (defining new constructs in terms of existing ones), deliberately excluding orthophrase (adding orthogonal features) and metaphrase (altering interpretation rules). This ceiling is a deliberate trade-off: it exists precisely to bound the attack surface. Agents requiring these patterns must implement them in application logic outside the CBCL layer.

\textbf{Trust infrastructure.} CBCL specifies algorithm-agnostic dialect signing (via an abstract \texttt{Signer} interface) but does not define key distribution, certificate authorities, or revocation. These are left to deployment infrastructure.

\textbf{Dialect identity and naming.} The DCFL preservation proof (\texttt{dcfl\_preserved}) requires unique dialect names per agent (\texttt{namesUnique}), since the \texttt{lang} tag dispatches by name. The true identity of a dialect is the content hash of its canonical encoding (RFC~9804); two dialects with the same name but different hashes are rejected at installation, mitigating name collisions and downgrade attempts. Dialect churn is bounded by R2 resource limits but requires deployment-level rate limiting for full mitigation.

\textbf{Dialect quality and bloat.} CBCL constrains dialect \emph{safety} but not dialect \emph{quality}. Nothing prevents propagation of redundant or poorly designed dialects. Agents can uninstall unhelpful dialects, but the protocol does not define curation mechanisms, leaving lifecycle management to agent policy.

\textbf{Performance.} The fuel-based termination model in the Lean formalization is sound but conservative. The reference implementation's runtime resource enforcement adds overhead proportional to the number of expansion steps.

\section{Related Work}

\subsection{Language-Theoretic Security}

Sassaman et al.~\cite{sassaman2013} established the foundation for language-theoretic security, observing that mismatch between an input language's complexity and the parser's recognizer class raises the likelihood of exploitable errors, and that unanticipated computational artifacts (``weird machines'') arise when input handling violates designer assumptions. Kaminsky, Patterson, and Sassaman~\cite{kaminsky2010} demonstrated this concretely with parser differential attacks against the X.509 PKI infrastructure. Sassaman et al.~\cite{sassaman2011} extended the analysis to network stacks, framing protocol insecurity as instances of the halting problem. Momot et al.~\cite{momot2016} developed a taxonomy of LangSec errors and proposed actionable CWE refactorings to address them. CBCL is, to our knowledge, the first protocol designed from the ground up on these principles for \emph{agent communication}, and the first to demonstrate that LangSec constraints are compatible with runtime language self-extension.

Von Hippel and Miyazono~\cite{vonhippel2025} argue that AI security is fundamentally a LangSec problem, identifying structured output parsing and tool-use capabilities in LLM-based systems as key attack surfaces. Their analysis reinforces the premise underlying CBCL: that securing agent communication requires constraining the computational complexity of the input language, not merely adding ad-hoc validation layers.

Fazeldehkordi et al.~\cite{fazeldehkordi2020} demonstrated that unconstrained communication patterns between distributed financial agents can enable DDoS through call-based flooding cycles, using static analysis to detect these patterns at design time. CBCL addresses a related concern at the protocol layer: R2 resource bounds ensure that processing any message, including dialect expansion, terminates within declared limits.

\subsection{Agent Communication Languages}

KQML~\cite{finin1994} introduced speech-act performatives for agent communication; FIPA-ACL~\cite{fipa2002} later codified 22 performatives with mental-state semantics. Both supported extensibility: KQML through an explicitly open performative set (communities could define new performatives by agreement), FIPA-ACL through \texttt{X-} prefixed user-defined message parameters. Gruber's Ontolingua~\cite{gruber1993translation} addressed the content layer separately. Building on McCarthy and Hayes's~\cite{mccarthyhayes1969} principle that what is represented can be specified independently of how it is computed, Gruber showed that interoperability between heterogeneous agents requires formal \emph{ontological commitment}: each agent must commit to the declarative specification of shared concepts (classes, relations, and the axioms constraining their use), with no commitment to the form or content of knowledge internal to the agent. CBCL's dialect mechanism inherits this commitment architecture; a dialect definition is a declarative specification to which the installing agent publicly binds itself, independent of its internal representation. Gruber left the process of reaching consensus on shared ontologies as an open problem, and treated ontology specification as separable from the communication protocol, so that extending shared meaning required out-of-band agreement invisible to protocol-level safety analysis. CBCL addresses the consensus problem directly: agents propose, query, and adopt dialect definitions as ordinary protocol messages. Singh~\cite{singh1998} identified a complementary flaw: reliance on unverifiable mental states rather than observable social commitments. CBCL makes dialect installation a protocol-level action subject to machine-checked safety constraints.

Yang et al.~\cite{yang2025} survey AI agent protocols across seven evaluation dimensions. Deng et al.~\cite{deng2025} survey security threats to AI agents, organising them around prompt injection, tool-use risks, and adversarial multi-agent interactions. CBCL's DCFL constraint directly addresses the parser-level attack surface underlying prompt injection: because dialect extensions are structurally distinct from message content, an agent's parser never conflates instructions with data. Zhang et al.~\cite{zhang2026clawworm} demonstrate this concretely: their ClawWorm worm achieves a 64.5\% attack success rate against the OpenClaw framework by exploiting a \emph{flat context trust model} in which LLMs cannot distinguish owner instructions from arbitrary channel input. CBCL's \texttt{lang}-scoped dialect mechanism, combined with R1--R3 verification, addresses this root cause by making provenance syntactically explicit. Zhou et al.~\cite{zhou2025} show that projecting high-dimensional LLM states to natural-language tokens is many-to-one and non-invertible, compounding information loss across conversational turns. A decidable formal language avoids this lossy bottleneck by making coordination semantics explicit in the message structure. Borazjanizadeh and Piantadosi~\cite{borazjanizadeh2024} provide complementary evidence. Pairing LLMs with a formal symbolic engine (Prolog) outperforms pure natural-language chain-of-thought on deductive tasks, suggesting that structured representations improve multi-step reasoning.

\subsection{Language Extensibility and Formal Methods}

McCarthy's 1982 proposal~\cite{mccarthy1982} envisioned open-ended business communication. Flatt~\cite{flatt2012} showed how language extensibility can be systematized through first-class language definitions in Racket. CBCL adapts this principle for adversarial distributed environments and makes the safety constraints explicit.

Demers et al.~\cite{demers1987} established the $O(\log n)$ convergence bound for epidemic information dissemination, which underlies CBCL's dialect propagation protocol. Jelasity et al.~\cite{jelasity2005} developed theoretical foundations for gossip-based aggregation, proving that the per-round convergence factor is independent of network size; their gossip framework informs CBCL's peer-selection strategy. RFC~9804~\cite{rfc9804} specifies a canonical form for S-expressions used in CBCL for deterministic signature verification.

\section{Conclusion}

CBCL demonstrates that the LangSec principle of matching input language complexity to recognizer capability can be applied even to self-extending protocols. By constraining dialect definitions to declarative pattern-template substitutions within S-expression syntax, we show that meaningful runtime extensibility and provable parsing safety are not in conflict. Because safety constraints are machine-checked at installation, agents can create and adopt dialects ad hoc, without external coordination.

As autonomous agent systems proliferate, the choice of communication substrate becomes a security decision. Natural language and unrestricted schemas offer maximal flexibility but cannot guarantee that two agents interpret the same message identically, a source of coordination failure even among cooperative agents and a fundamentally unsecurable attack surface in adversarial settings. CBCL offers an alternative: a formally specified, machine-verified protocol where every message, including those that extend the language itself, is parsed in bounded time by a deterministic pushdown automaton. If agents communicate in languages whose validity is undecidable, no monitor can reliably determine what they have agreed to do. Formally bounding what agents can express to each other is a precondition for oversight.

\smallskip\noindent\textbf{Availability.} The reference implementation, Lean~4 formalization, verified parser binary, and IETF Internet-Draft are available at \url{https://codeberg.org/anuna/cbcl-rs} under the Apache~2.0 license. The Nostr transport binding is at \url{https://codeberg.org/anuna/cbcl-nostr}.

\section*{Acknowledgments}

Thanks to Claire Barnes, David Factor, Mat Mytka, Mark Pesce, Marc Ahrens, Prof.\ Ingo Weber, Dr Max Ott, Dr Mark Staples, Dr Ho-Pun Lam, Dr Adnene Guabtni, and DZJ for helpful discussions. Thanks to Ric Richardson and the Office for Innovation for championing this effort. This work was inspired by problems encountered during a Science and Industry Endowment Fund project through CSIRO's Data61, and the need for ad-hoc knowledge sharing across agricultural supply networks.

\smallskip\noindent\textbf{AI Disclosure.} In accordance with IEEE policy, the author discloses that AI assistants were used throughout this work, including drafting and editing this manuscript, assisting with the Lean~4 formalization, contributing to the reference implementation, and aiding in literature review. Tools used include llm-md, Anthropic's Claude Code, and OpenAI's Codex. Agent orchestration was coordinated using hence, the author's defeasible-logic meta-control planner for multi-agent LLM task coordination. All work was directed, reviewed, and validated by the author. The author takes full responsibility for the content of this publication.


\bibliographystyle{IEEEtran}

\begingroup
\sloppy
\hbadness=10000

\endgroup

\end{document}